\documentclass{appolb}
\usepackage{epsfig}
\usepackage{amssymb}

\begin{document}
\title{Neutrino Astronomy with ANTARES
\thanks{Presented at XXXIV Int. Symposium on Multiparticle Dynamics, 
ISMD 2004, Sonoma County, CA, July 30, 2004}%
}
\author{Teresa Montaruli
\address{Universit\'a di Bari and INFN}\\
on behalf of the ANTARES Collaboration
\address{http://antares.in2p3.fr}
}
\maketitle
\begin{abstract}
ANTARES is a project aiming at the operation of an
underwater detector at a depth of 2.5 km close to Toulon in the 
South of France. The detector is expected to be completed at the beginning
of 2007.
The main purpose of the experiment is the detection of 
high energy neutrinos produced in astrophysical sources.
Being weakly interacting, neutrinos could potentially be more
powerful messengers of the universe compared to photons, but
their detection is challenging.
The technique employs phototubes to detect the arrival time and
the amplitude of photons emitted by neutrino charged secondaries 
due to the Cherenkov effect.
ANTARES will contribute significantly in the field of neutrino
astronomy, observing the Galactic Centre with
unprecedented pointing capabilities.
\end{abstract}
\PACS{PACS numbers come here}

\section{Introduction: neutrino astronomy and telescopes}
\label{intro}

Neutrinos are considered possible new messengers of the universe
since their properties differ from those of photons, that
currently provide most information.
They could improve our understanding of the sources and mechanisms capable
of accelerating cosmic rays up to energies larger than $10^{20}$ eV, that are
observed by extensive air shower arrays.
Photons of energies larger than 10 TeV cannot bring information from 
distances $\gtrsim 100$~Mpc since they interact by pair production with 
background photons. 

A deep connection between high energy $\gamma$ emissions and
$\nu$ production exists according to the model of the
``beam dump''. Protons or nuclei
accelerated by an engine, interact on a gas of matter or photons. 
These interactions result mainly in neutral and charged pions
which decay into photons and neutrinos. This explains the expectation
that, if $\gamma$ absorption could be ignored, observed $\gamma$ fluxes and 
$\nu$ ones should have almost the same normalization and  
spectral shape. This should be a power law with spectral index 
around $-(2\div 2.5)$, as expected from Fermi acceleration processes. 

Gamma-ray experiments are currently providing interesting results and
a new era for the observations above 1 TeV has been recently inaugurated by 
the HESS experiment \cite{HESS}. The experiment observed TeV photons from
the Galactic Centre with 9 standard deviations in only
12 h of observation, reconstructing a spectral index of the differential
energy spectrum above 165 GeV of $-2.21 \pm 0.09 \pm 0.15$, in disagreement 
with CANGAROO \cite{CANGAROO_cen}.
This region is extremely complicated due to the presence of Sgr
A$^*$, most probably a super-massive black hole, and of the
supernova remnant (SNR) Sgr A East. The measured spectrum could be interpreted
as due to purely electromagnetic processes, but also to proton acceleration
and photo-meson 
interactions with ambient radiation or interactions 
with the plasma close to Sgr A$^*$ \cite{Aharonian}. 
Both these last two scenarios imply high energy $\nu$ production. 

From about 18 h of data, HESS estimated a similar spectral
dependence than for the Galactic Centre for the SNR RX J1713-3946. The 
observed photon spectrum by CANGAROO was pointed out to be explained by
$\pi^{0}$ decay and not by electromagnetic processes \cite{CANGAROO}, 
though the result has been considered
controversial since the predicted spectrum seems to
exceed the EGRET observed flux from the nearby region. 
The spectrum measured by HESS does not show the cut-off in the 
CANGAROO spectrum and it is compatible with EGRET observations.
If the $\nu$ flux were equal to the one measured by HESS, 
event rates expected in ANTARES would be $<1$ event/yr.
Nevertheless, it is clear that the Galactic Centre 
region is getting more and more interesting for what concerns the
possibility of proton acceleration. The observation of this region can be 
done only from the upper hemisphere using $\nu$-induced muons, 
that is the topology with the best angular resolution in $\nu$
telescopes. 
Preliminary studies show that ANTARES will perform this observation 
with unprecedented
angular resolution, comparable to recent and forthcoming $\gamma$ experiments,
such as HESS, MAGIC, INTEGRAL, SWIFT and GLAST. These experiments
will provide alerts in case of bursting sources,
such as gamma-ray bursters (GRBs), and clues to select 
interesting candidates in order to enhance ANTARES sensitivity.

Other galactic sources, such as fast rotating neutron stars
in SNRs, plerions, associations of stars, 
and magnetars, are foreseen to predict up to
hundreds of muon neutrino 
events in a cubic-kilometer detector and a few events in
ANTARES (for a review see \cite{Burgio} and references therein). 
ANTARES effective area for muons reaches 0.05~km$^2$ for 
$E_{\nu} \gtrsim 10^6$~GeV.
The most promising models are already constrained 
by the AMANDA-II limits \cite{AMANDA, AMANDAnu}. This is the case of the
micro-quasar model in \cite{Distefano} for what concerns the persistent
source SS433, though this peculiar source is surrounded by a nebula
and hence the model could be particularly uncertain.

Extragalactic sources, such as active galactic nuclei and GRBs,
if transparent to nucleons, can produce event rates up to 
from a few to a few hundreds 
of events in a km$^3$ telescope.
As a matter of fact, if the neutrons can escape sources and produce
the UHE protons observed by extensive air shower arrays 
above $10^{18}$ eV, an upper limit can be derived as firstly shown in
\cite{WB} (hereafter W\&B limit).  
Nevertheless scenarios exist that evade the limit by
considering other proton spectral dependences than $E^{-2}$
and sources that are not transparent to nucleons (see 
Ref.~\cite{MPR}, hereafter MPR limit) . 
AMANDA-II \cite{AMANDA,AMANDAnu} is already at the level
of testing the limit for completely opaque sources in \cite{MPR}, while
the W\&B flux limit will be tested after 1 yr of operation of the 
km$^3$ detector IceCube \cite{IceCube}.

It is probable that neutrino detection will allow also to
address the problem of the observation of the
ultra-high energy cosmic rays (CRs), since these 
could be produced by the same processes in sources.
Other granted high energy $\nu$ sources should be 
CR interactions with the cosmic microwave background and 
with the galactic interstellar matter (for the former
event rates in ANTARES are of the order of a few events per year).
Moreover, speculative top-down production processes can be envisaged,
where supermassive particles or topological defects decay 
into neutrinos. 

Neutrino telescopes detect the Cherenkov light emitted by 
charged particles produced in $\nu$ interactions by 3-D arrays of 
optical modules (OMs). OMs are pressure resistant glass spheres containing 
phototubes (PMTs), located in polar ice or sea/lake water depths in order 
to reduce the surface $\mu$ flux by orders of magnitude. 
From the times of PMTs hit by the Cherenkov light, tracks can be
reconstructed, while the amplitudes allow an energy reconstruction.
Neutrino telescopes were originally optimized to detect upward-going muons from
$\nu_{\mu}$ charged current (CC) interactions. Their direction of flight allows
to discriminate atmospheric muons, since $\nu$s are the only 
atmospheric shower secondaries capable of crossing the Earth.
Since the $\nu$ cross-sections and the $\mu$ range increase with energy,
the effective target mass for $\nu$ interactions becomes larger with
energy. As a matter of 
fact, the larger the energy the better the performances of this technique, 
up to a saturation that depends on the detector dimensions. Moreover,
for $E_{\nu} \gtrsim 10$ TeV the muon has the same direction of the parent 
neutrino allowing to point sources.

In view of recent results on neutrino oscillations, 
it is now considered important to detect other flavor neutrinos.
Flavor ratios at sources from meson decays, in the hypothesis that
the environment density is such that 
all muons decay, are
$\phi_{\nu_{e}} : \phi_{\nu_{\mu}} : \phi_{\nu_{\tau}} = 1 : 2 : 
\lesssim 10^{-5}$~\cite{Bugaev}.
In light of solar and atmospheric $\nu$ results and of
constraints from reactor experiments, oscillations through baselines 
larger than tens of kpc would lead 
to $\phi_{\nu_{e}} : \phi_{\nu_{\mu}} : \phi_{\nu_{\tau}} = 1 : 1 : 1$.
Since the $\nu$ cross section increases with energy, 
CC interactions during propagation through the Earth
prevent $\nu_{\mu}$ and $\nu_{e}$ from reaching the detector
(unless for $\nu_{\mu}$ the muon range is 
larger than the distance of the $\nu$ vertex to the
detector). 
The Earth shadowing becomes important above 1 PeV and neutrinos can reach the
detector only from the horizon or from the upper hemisphere. 
Hence for UHE $\nu$ detection it is necessary to have good
efficiencies in reconstructing horizontal events, to use
energy estimators to suppress the atmospheric $\mu$ background, and 
discriminate showering events inside the instrumented region from
crossing atmospheric $\mu$ tracks.
The case of tau $\nu$s is peculiar: they are never absorbed during propagation 
through the Earth, though they loose
energy, since they regenerate in $\tau$ decays \cite{Bugaev}.
They can produce background free topologies, such as the 'double bang' 
events, where the shower
from the $\nu$ vertex and the shower from the electronic and hadronic
$\tau$ decay channels are connected by a $\tau$ track,
long enough to separate the showers.
These interesting events, that would prove $\nu$ oscillations in
an astrophysical beam, are expected to be very rare, depending on the assumed
spectrum,
due to the fact that a $\tau$ track is $>50$~m for $E_{\tau} \gtrsim 1$~PeV).

\section{The ANTARES project}

\begin{figure}[htb]
\begin{center}
 \begin{tabular}{cc}
\epsfig{file=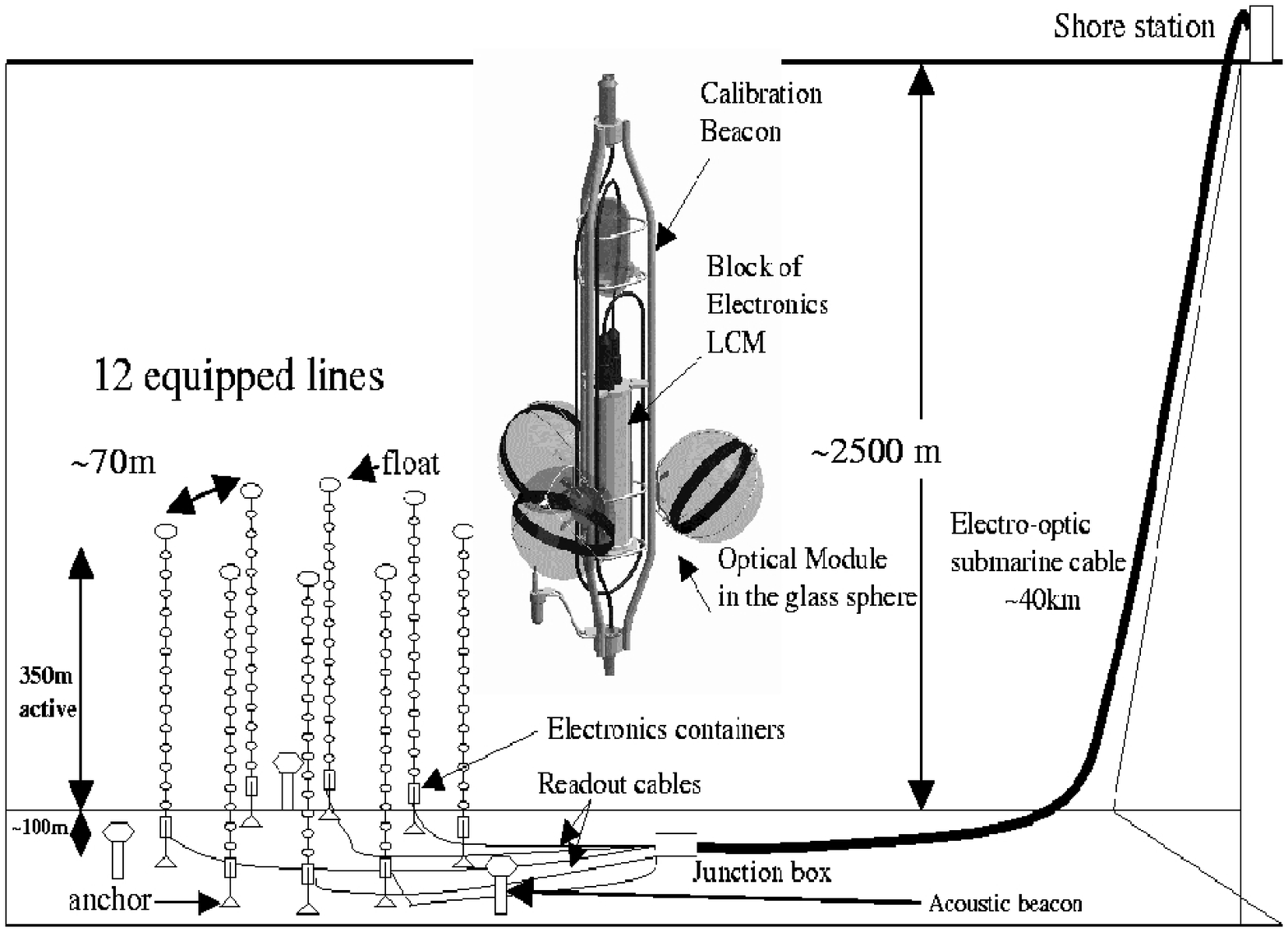,width=6.5cm,height=5.cm}&
\epsfig{file=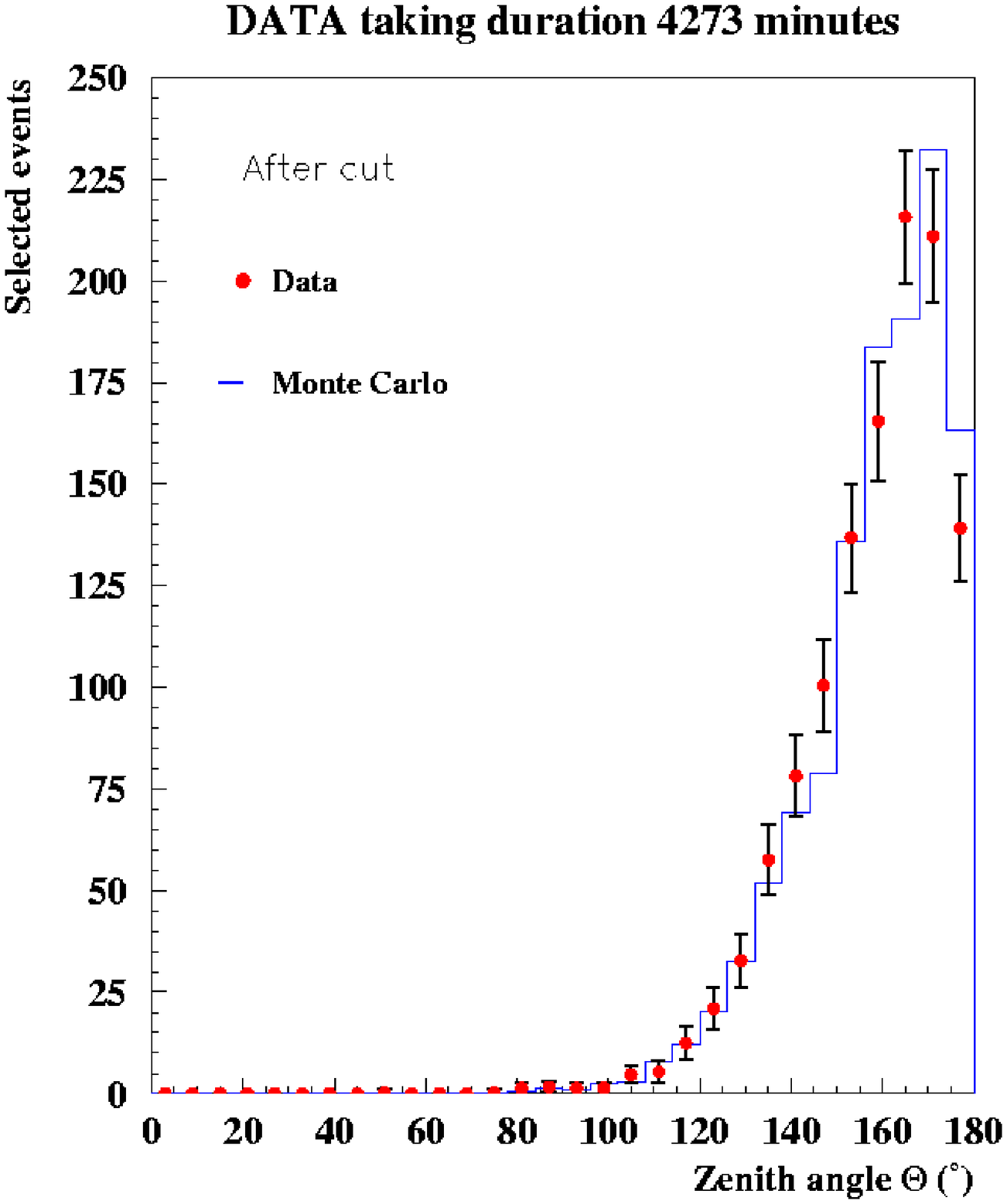,width=6.cm,height=5.cm}
\end{tabular}
\caption{\label{fig1} 
{\bf On the left:} Layout of the ANTARES detector. 
The detail of a storey is shown. {\bf On the right:}
reconstructed atmospheric $\mu$ angular distribution from Line 5 data.
The areas of the curves are normalized to each other.}
\end{center}
\end{figure}

The ANTARES Collaboration was formed in 1996 and comprises physicists, 
astronomers, engineers and sea experts from France, Germany, 
Italy, the Netherlands, Russia, Spain and The United Kingdom.
The installation of the detector has started in a 2.5 km deep site about
40 km off-shore Toulon ($42^{\circ}$50'N,$6^{\circ}$10'E).
The set-up of the telescope is shown in Fig.~\ref{fig1}. Twelve lines 
form an octagon with average distances of about 65~m. 
Being 450 m high, they are kept taut by buoys at the top end and by anchors
at the bottom. 
The first 100 m from the bottom of the line are not instrumented, 
so that OMs are far enough from seabed mud and to
have additional water volume where the Cherenkov light can be produced.
Each line is connected to a junction box (JB), distributing 
data and power from/to the detector from/to the shore station 
through the electro-optical cable (EOC).
Both the JB and the EOC are already installed and performing
well since Oct. 2001 and Dec. 2002, respectively.
Each line is instrumented with 25 storeys separated vertically 
by 14.5 m, holding 3 OMs, each containing 
a 10'' 14 stage Hamamatsu R7081-20 phototube (PMT). 
The 3 PMTs are mounted at $45^{\circ}$ from the vertical looking downwards. 
As a matter of
fact, the OM transparency reduction is negligible for surfaces exposed
below the horizon \cite{sedimentation}. 
OMs, that have to resist 260 bars during normal operation, 
are transparent in the 400-500 nm wave-length region. They
include a $\mu$-metal cage preventing the Earth magnetic field to
deflect electron paths \cite{OM} and a blue LED for time calibrations.
At the moment of writing (Oct. 2004), 250 OMs of the 900 have been
completed. 

The main specifications that ANTARES
PMTs must satisfy are: a transit time spread (TTS)
of less than 3 ns (FWHM), a peak to valley ratio larger than 2, 
a dark count of less than 20 kHz for a 0.25 photoelectron (pe) threshold,
and a gain larger than $5 \cdot 10^7$. Laser calibrations in a dark room
using a prototype line made of 5 storeys have shown that, after corrections
of clock delays between consecutive storeys, the achievable timing 
resolution is $\sigma \sim 0.9-1.2$ ns.
Each storey contains a local control module, a titanium cylinder containing 
the electronics. The digitization of the PMT signals is done by 
the Analogue Ring Samplers (ARS), ASIC 
full custom chips. Signals are processed according to a
single photoelectron mode, that provides the time stamp and charges, 
and a waveform mode for $\sim 2\%$ 
large amplitude or complex signals by waveform digitizers (1 GHz sampling). 
The motherboard for each PMT is equipped with 3 ARS and 2 of them are
activated in turn to reduce dead time. The digitized data will
be sent to shore through optical fibers at a
rate of $\sim 1$ GB/s for further processing with a PC farm. Online
filters, based on clustering algorithms that rely on time correlations between
hits through velocity of light in water, 
will reduce the amount of data by a factor of $\sim 10^3$. As a matter of fact,
this will allow the rejection of most of the background data due to
$^{40}$K decays and bioluminescence.
The data filter efficiency for upgoing $\nu$ events, 
under continuous improvements,
increases with energy and is of the order
of 50\% at 1 TeV and 70\% at at 10 PeV. It can be dramatically
improved, especially at low energies, when the directional
information is used. This is the case of GRBs for which ANTARES
will receive alerts from the GCN network of satellites and of
candidate sources that can be followed in dedicated periods.  
   
Since lines move in water, the position and orientation
of OMs must be monitored.
Relative positioning of storeys, which should be accurate
at the 0.5 ns level, is provided by a system of
acoustic beacons at seabed and hydrophones in storeys.
The orientation is determined through compasses and tiltmeters in 
the electronics container. The PMT TTS will be calibrated using the LEDs in
the OMs, an optical beacon each 5 storeys and a laser beacon at the bottom
of some of the strings.
The absolute timing will be $\sim 1$~ms accurate 
by correlating the time provided by the GPS system to the internal clock,
a 20 MHz high accuracy electrical signal generated on-shore and
converted into an optical signal distributed to the detector.

ANTARES has done many tests and sea campaigns for the
site qualification. This phase includes the deployment of a 
test line with 7 PMTs in Nov. 1999 (Line 5) at 1.2 km depth, 
that has reconstructed atmospheric muons requiring a 5-fold
coincidence. Their
zenith angle distribution is shown in Fig.~\ref{fig1} (on the right). The 
shape is in agreement with the Monte Carlo.
In 2003 a prototype line with 5 storeys and a 
reduced version of the final instrumentation line 
devoted to environmental parameter monitoring were 
operated for a few months and recovered.
Both lines were successfully installed within a few meters 
from their nominal positions and connected to the JB, proving the
feasibility of marine operations.  
Measurements of the heading of storeys
have shown that the line moves as a rigid body.
Beside the successful operation of these lines, a few problems occurred
and remedies have been found for the forthcoming lines. 
The failure in the clock transmission prevented
the data taking at nanosecond precision level. Nevertheless, the
counting rate monitor has allowed taking optical background data for a period 
of $\sim 100$ d. Large and short lived peaks of a few seconds due
to light emitting animals have been detected over a baseline of about 
60 kHz due to $^{40}K$ decay and bacteria. This baseline is variable and
can increase up to values of 200 kHz. Correlations of this background to
sea currents, e.g. the water rotation due to the Coriolis force,
and to storey heading have been demonstrated.

\subsection{Physics studies}

\begin{figure}[htb]
\begin{tabular}{cc}
\epsfig{file=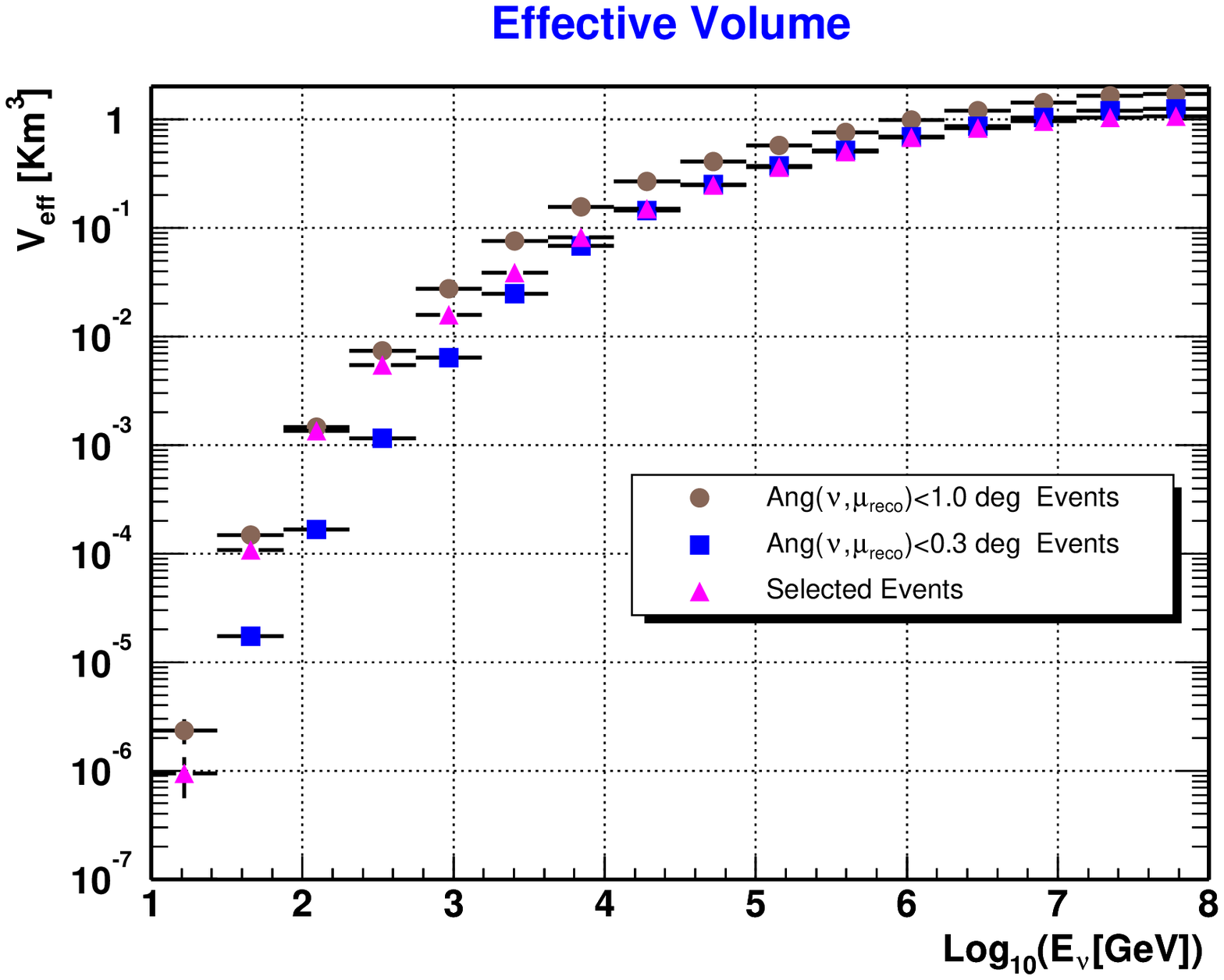,width=6.cm,height=6.cm}&
\epsfig{file=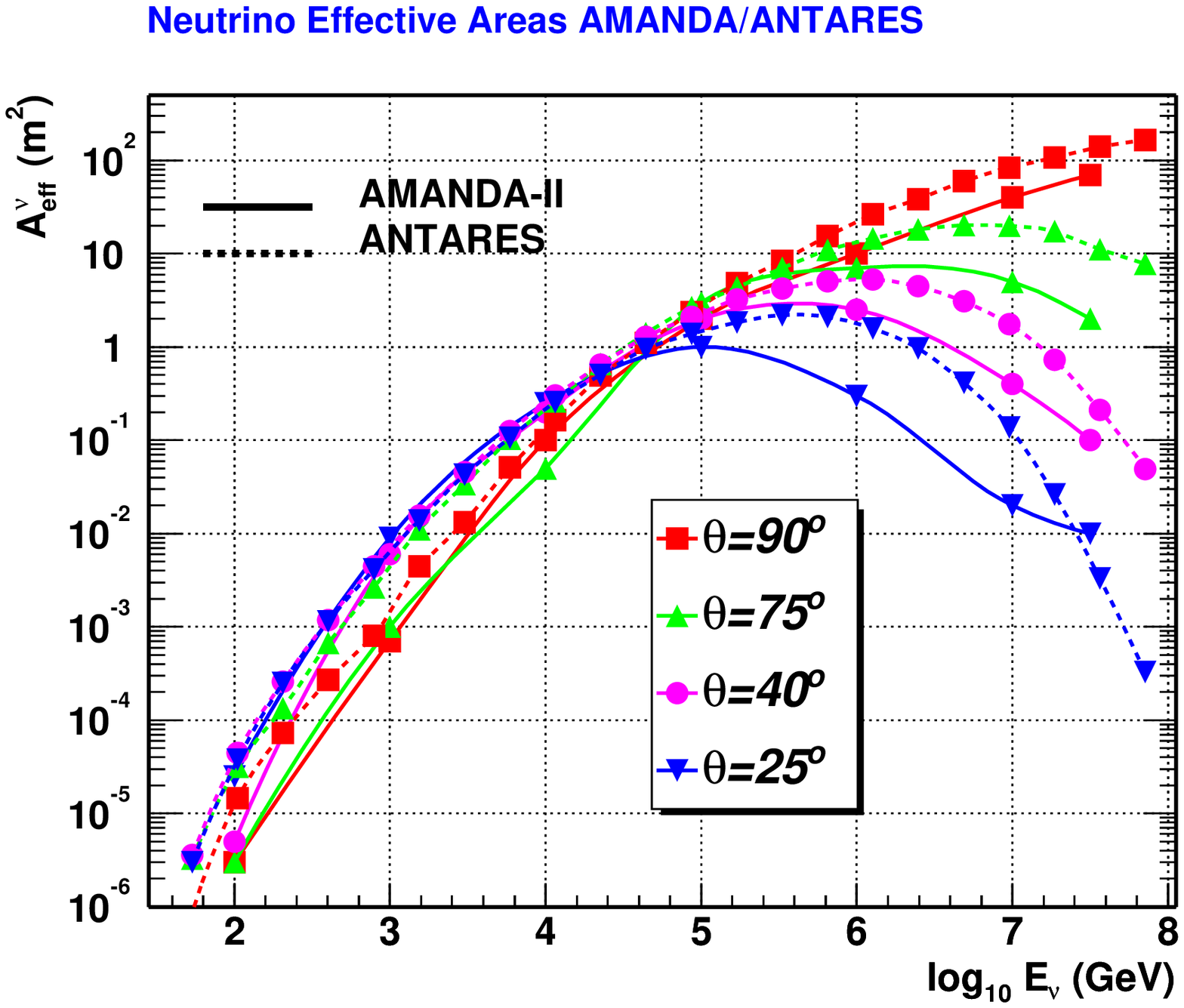,width=6.5cm,height=6.cm}
\end{tabular}
\caption{\label{fig2}  {\bf On the left:} the ANTARES effective volume for 
$\nu_{\mu}+\bar{\nu}_{\mu}$ vs $\nu$ energy. The triangles 
indicate the volume after quality requirements are applied to the
reconstruction algorithm. In order to understand the pointing capabilities
after this cuts, the squares indicate the effective volume
that would be obtained when the angle between the simulated
$\nu$ and the reconstructed $\mu$  is $\le 0.3^{\circ}$. While at low
energy the difference between the two curves is due to the kinematic
angle of the $\nu-\mu$ interaction, at energies $\gtrsim 10$ TeV 
it is dominated by the intrinsic angular resolution.
For searches where also time constraints can be used to reject the
atmospheric muon background looser cuts can be applied, corresponding
to a larger $A^{eff}$ (dots for the $1^{\circ}$ cut).
{\bf On the right:} Comparison of the ANTARES neutrino effective area (dashed
lines) with the AMANDA-II one \protect\cite{AMANDA} (solid lines)
for zenith angles.
The decrease of the area in the vertical region (inverted triangles) 
compared to the horizontal one
(squares) at energies $\gtrsim 100$~TeV is due to the shadowing effect 
of the Earth due to the cross-section increase with energy.}
\end{figure}

The parameters that best describe neutrino telescope
performances are the effective volume and the neutrino effective area,
that are deeply connected (see Fig.~\ref{fig2}).
They need to be expressed as a function of neutrino energy
in order to have a unique definition of the energy for
different experiments and different topologies of events. 
As a matter of fact,
secondaries such as muons from $\nu_{\mu}$ CC interactions 
loose energy during propagation outside and in the instrumented region,
hence their energy cannot be uniquely defined.
The effective volume of neutrinos is the volume of a 100\% efficient
detector for observing $\nu$ interactions that would obtain the same
event rate as ANTARES for a given interaction rate (that depends on the
$\nu$ flux, the target medium and the cross-section): $V_{eff} = 
\epsilon V_{gen}$, where $V_{gen}$ is the volume where neutrino CC
interactions can produce secondaries than can emit detectable 
Cherenkov light (for $\nu_{\mu}$ it depends on the muon range). 
This definition includes the tracking and
selection requirements: $\epsilon = \frac{N_{sel}(E_{\nu},\Omega)}
{N_{gen}(E_{\nu},\Omega)}$, where $N_{gen}$ is the number
of generated events in the generation volume $V_{gen}$ and
$N_{sel}$ is the number of selected events.
The $\nu$ effective area 
is the sensitive area 'seen' by $\nu$'s producing detectable $\mu$'s
when entering the Earth:
\begin{equation}
A^{eff}_{\nu}(E_{\nu}, \Omega) = V_{eff} \cdot
N_{A} \rho \sigma_{\nu}(E_{\nu}) \cdot P_{Earth}(E_{\nu}, \Omega) \, ,
\label{eq:aeff}
\end{equation}
\noindent 
where $P_{Earth}$ is the absorption probability through the Earth,
and $1/(N_{A} \rho \sigma_{\nu})$ is the interaction length in the
matter of the generation volume.
This is a function of the energy and of the local
angles and it allows the calculation of event rates
for a $\nu$ model predicting
a spectrum $\frac{d\Phi}{dE_{\nu}d\Omega_{\nu}}$: 
\begin{equation}
N_{\mu} = \int\int dE_{\nu}d\Omega_{\nu} 
A^{eff}_{\nu}(E_{\nu}, \Omega_{\nu})
\frac{d\Phi}{dE_{\nu}d\Omega_{\nu}} \, .
\end{equation}
Given this definition, it is understandable why the
dimension of this quantity is orders of magnitude
lower than the geometrical dimensions of the array. Simply
such huge detectors are reduced to areas of the order of 
tens of m$^2$, due to the weakly interacting
properties of neutrinos. 
The area being strongly energy dependent,
detectors respond in different regions to $\nu$'s with different spectra:
the harder the spectrum the highest is the mean energy of the
corresponding detectable events (e.g. $\sim 100$~GeV
for typical $E^{-3.6}$ atmospheric $\nu$ spectra, 
$\sim 10$~TeV for typical $E^{-2}$ cosmic $\nu$ spectra). 

\begin{figure}[htb]
\begin{tabular}{cc}
\epsfig{file=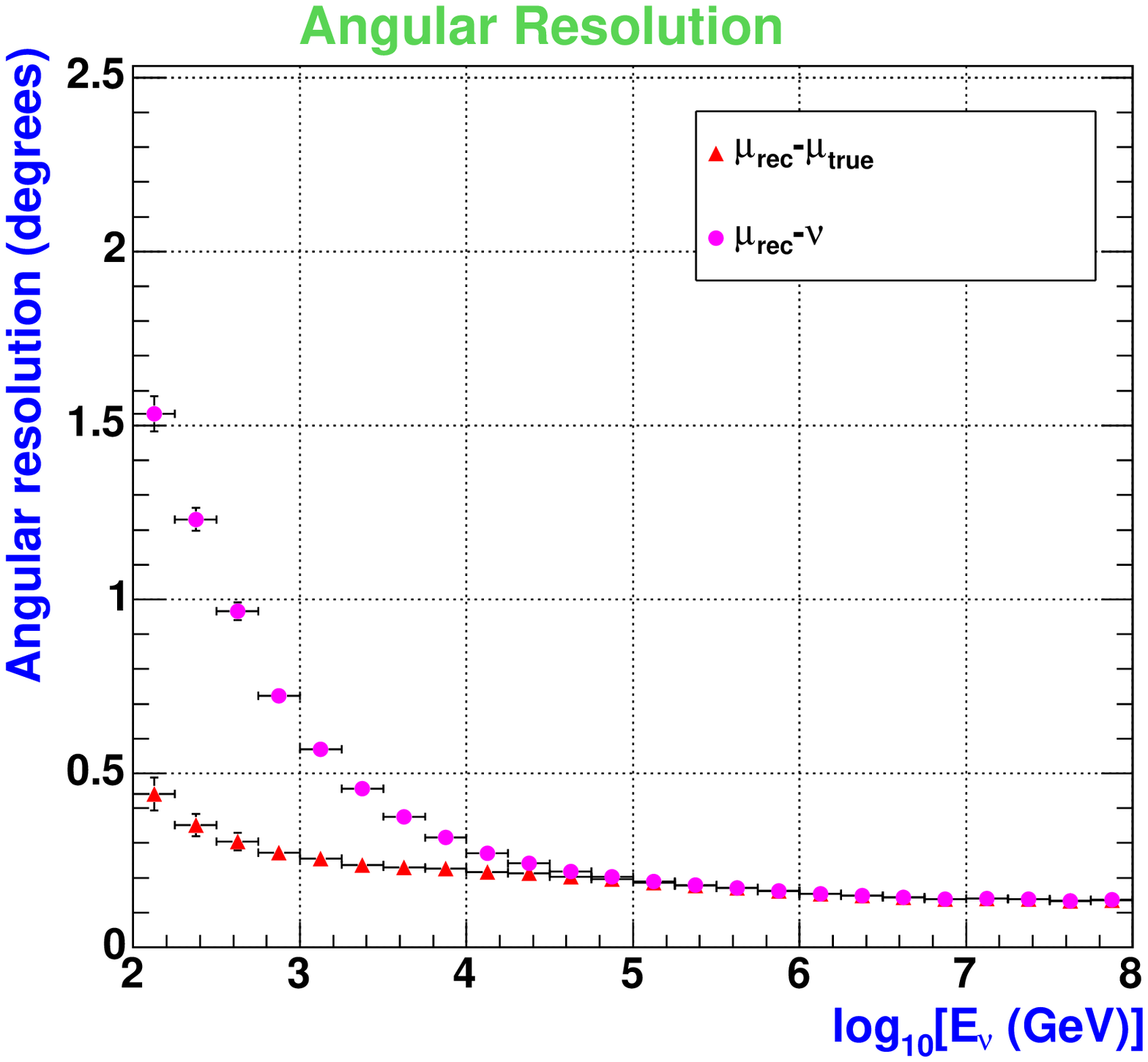,width=6.cm,height=6.cm}&
\epsfig{file=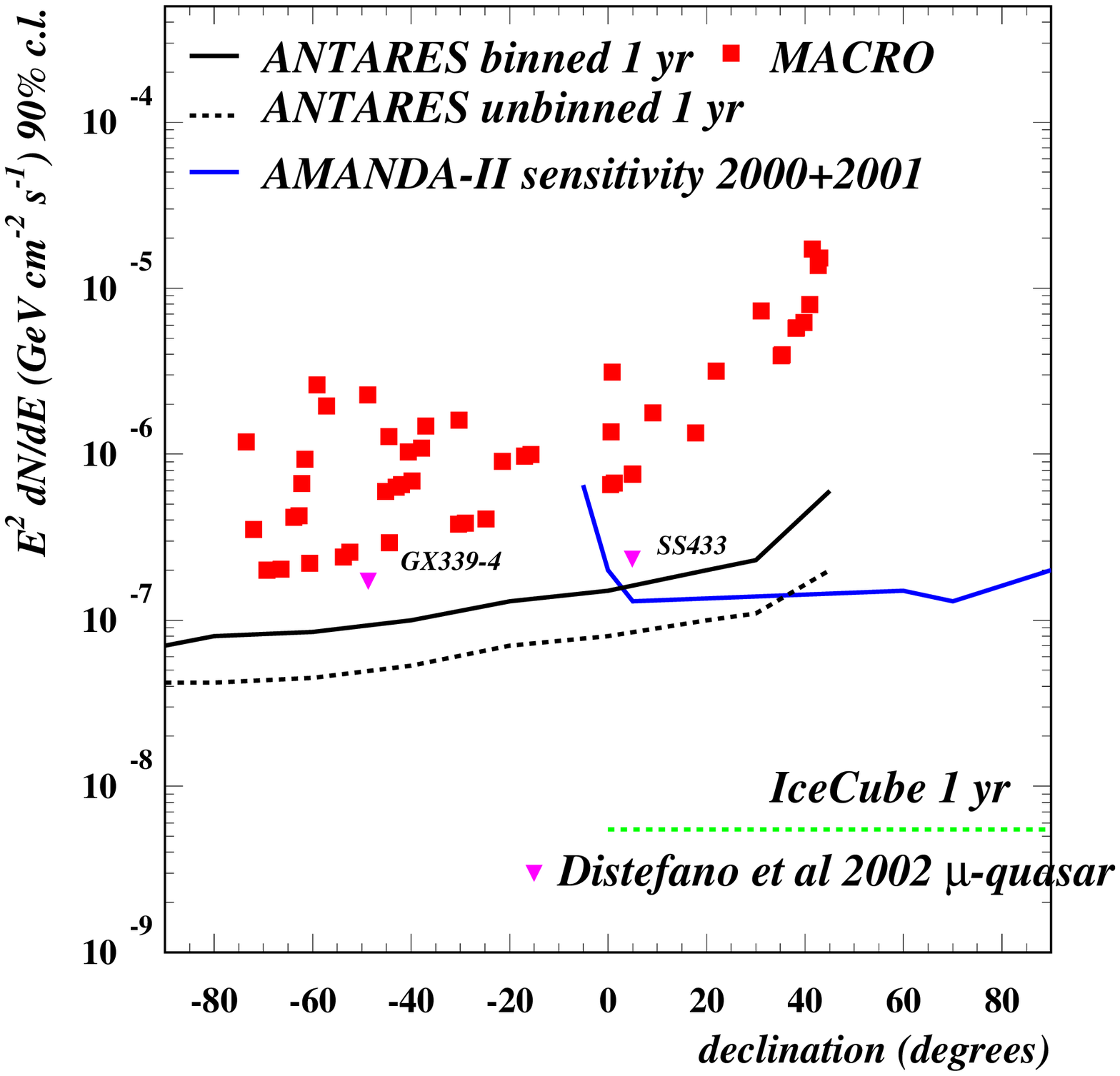,width=6.5cm,height=6.cm}
\end{tabular}
\caption{\label{fig3}
{\bf On the left:} ANTARES expected angular resolution vs $\nu$ energy. The
dots represent the median angle between
the simulated $\nu$ and the reconstructed $\mu$ direction
(that is the pointing capabilities for a neutrino source), while the
triangles are the median of the angle between the simulated
muon and the reconstructed one (the intrinsic angular resolution).
{\bf On the right:} Upper limits (90\% c.l.) on $E^{-2}$ neutrino fluxes 
as a function of the source declination for MACRO (squares) \cite{MACRO},
expected sensitivity of AMANDA-II corresponding to 2000-1 data
\cite{AMANDAnu}, IceCube \cite{IceCube} and ANTARES \cite{ANTARESpoint}
(for a search method using a grid in the sky and an unbinned method
based on likelihood ratio). AMANDA and IceCube lines cover the
upper hemisphere region since upgoing $\nu$ events are used. The triangles
indicate the expected $\nu$ flux from two persistent micro-quasars 
as calculated in \cite{Distefano}. The flux for SS433 is excluded by AMANDA and
after 1 yr of ANTARES data.}
\end{figure}

\begin{figure}[htb]
\begin{center}
 \epsfig{file=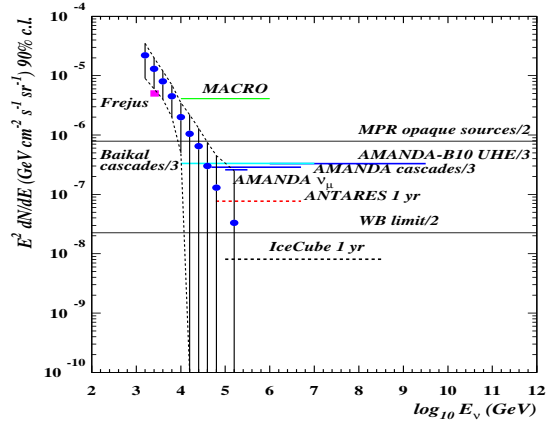,width=8.cm,height=6.cm}
\end{center}
\caption{\label{fig4} 
90\% c.l. limits on diffuse $E^{-2}$ fluxes of $\nu_{\mu}+\bar{\nu}_{\mu}$
in the hypothesis of $\nu$ oscillations
as measured by AMANDA-II \cite{AMANDAnu}, Baikal 
\cite{Baikal} 
and MACRO \cite{MACRO1}. Limits for other flavors than $\nu_{\mu}$ 
(cascades) have
been divided by the number of contributing flavors. 
Also the expected sensitivities for ANTARES 
and IceCube \cite{IceCube} are shown.
Dots are the measured atmospheric $\nu$ flux by AMANDA-II 
\protect\cite{AMANDA}.}
\end{figure}

Another fundamental parameter for
point-like source searches is the angular resolution for tracks
$\Delta\theta$, since
the signal to noise ratio is given by:
\begin{equation}
S/\sqrt{N} \propto \sqrt{AT}/\Delta\theta \, ,
\end{equation} where
$A$ is the effective area and T the detection time.
The angular resolution as obtained by ANTARES simulations is
shown in Fig.~\ref{fig3} (on the left). The plot shows 
the median angle between
the $\nu$ source and the reconstructed muon and the 'intrinsic 
angular resolution', that is the angle between the 'true' muon and
the reconstructed one. Its limiting value is  
$\sim 0.2^{\circ}$ for $E_{\nu}\gtrsim 10$ TeV. 
The sensitivity of ANTARES for 1 year of data taking as a function of 
declination for two different methods of searches for point-like sources
is shown in Fig.~\ref{fig3} (on the right), where it is compared
to other experiments. 
Studies on the sensitivity of ANTARES to a diffuse muon neutrino flux from
populations of sources have brought to the result given 
in Fig.~\ref{fig4}, where
it is compared to other experiments.
In this case the best estimator of the energy between those tested has
resulted to be the number of hit PMTs, that is used to make an high energy
cut to reject atmospheric muons and neutrinos. This possibility is due to
the steeper spectrum of atmospheric fluxes compared to fluxes from sources.  
\section{Summary and Outlook}
ANTARES will observe the Southern hemisphere sky using neutrinos.
It will complement the detectors at the South Pole and represents
an important step to acquire the proper know-how to deploy 
a km$^3$ kilometer detector in the Mediterranean.
The first line of the detector is planned to be deployed and connected to
the junction box in the summer of 2005. 
In the meanwhile, two more lines will be deployed. The MILOM combines 
a storey with optical modules, calibration devices (laser, LED beacons and
acoustic positioning modules), and instruments for the monitoring 
of optical and mechanical environment of the site. The so-called Line zero, 
with 25
storeys but no PMTs inside glass spheres, will allow to realistically test
the electro-mechanical cables and all the containers.
The assembly and test of both lines will validate the final line
integration procedure.

\end{document}